\documentclass[onecolumn,
]
{revtex4}
\usepackage{float}
\usepackage{amssymb}
\usepackage{epsfig}
\newcommand{\beq}{\begin{eqnarray}}
\newcommand{\eeq}{\end{eqnarray}}
\newcommand{\bege}{\begin{equation}}
\newcommand{\enge}{\end{equation}}
\newcommand{\noi}{\noindent}

\begin{document}

\title{Conformal Klein-Gordon equations and quasinormal modes}

\author{R. da Rocha}
\email{roldao@ifi.unicamp.br}
\affiliation{DRCC - Instituto de F\'{\i}sica Gleb Wataghin, Universidade Estadual de Campinas,
CP 6165, 13083-970\\ Campinas, SP, Brazil.}

\author{E. Capelas de Oliveira}
\email{capelas@ime.unicamp.br}
\affiliation{Departamento de Matem\'atica Aplicada, IMECC, Unicamp, CP6065, 13083-859, Campinas, SP, Brazil.}


\begin{abstract}
Using conformal coordinates associated with conformal 
relativity -- associated with de Sitter spacetime homeomorphic projection into Minkowski spacetime -- we obtain a conformal Klein-Gordon partial differential 
equation, which is intimately related to the production of quasi-normal modes (QNMs) oscillations, in the context 
of electromagnetic and/or gravitational perturbations around, e.g., black holes. 
While QNMs arise as the solution of a wave-like equation with a P\"oschl-Teller potential,
here we deduce and analytically solve a conformal `radial'
d'Alembert-like equation, from which we \emph{derive} QNMs formal solutions, in a proposed alternative to more completely describe
QNMs. As a by-product we show that this `radial' equation 
can be identified with a Schr\"odinger-like equation in which the potential is exactly 
the second P\"oschl-Teller potential, and it can shed some new light on the investigations 
concerning QNMs.

\end{abstract}
\maketitle
\section*{Introduction}

Quasi-normal modes (QNM) arise in the context of general relativity as electromagnetic or gravitational perturbations 
occurring in the neighborhood of, e.g., Schwarzschild, Kerr, Reisnerr-Nordstr\o m \cite{15}, and Kerr-Newman spacetimes, and their
 investigation in the Schwarzschild background was started in \cite{12,14}. 
It is well known that no normal mode oscillations is produced in the process of 
emission of gravitational waves, but only quasi-normal oscillation modes, representing a oscillatory damped wave
 \cite{1,2,10,11,14,16,19} while, as yet nothing is known for nonlinear stellar oscillations in general relativity 
 \cite{161} and in the collapse of a star to form a black hole \cite{14}. 
As defined in, e.g.  \cite{3,11}, QNMs are the eigenmodes of the homogeneous wave equations, describing these perturbations --
 with the boundary conditions corresponding to outgoing waves at the spatial infinity and incoming waves at the horizon. 
The paramount interests to QNMs have been mainly introduced by \cite{18,19}. QNMs can bring imprints of black holes and can be detected 
in the gravitational wave framework \cite{11}.

In one hand, QNMs equations can be derived if we  introduce a projective approach, 
 namely the theory of hyperspherical universes, developed by Arcidiacono\cite{arci1} several years ago and, more specifically, in the so-called conformal case.
When we write Maxwell equations in six dimensions, with six projective coordinates (we have, in these coordinates, a Pythagorean metric) a natural problem arises, namely, to provide a physical version of the formalism, i.e. to ascribe a physical meaning to the coordinates. For  this theory, there are two possible different physical interpretations: a bitemporal interpretation and a biprojective interpretation. In the first case (bitemporal) we introduce a new universal constant $c'$ and the coordinate $x_5 = i c't'$ where $t'$ is interpreted as a second time; we thus obtain in cosmic scale the so-called multitemporal relativity, proposed by Kalitzen\cite{kali}. The set of Maxwell equations obtained in this theory generalizes the equations of the unitary theory of electromagnetism and gravitation, as proposed by Corben\cite{corb}.

On the other hand (our second, biprojective case) we can interpret the extra coordinate, 
$x_5$, as a second projective coordinate. We then obtain the so-called conformal 
projective relativity, proposed by Arcidiacono\cite{arci2}, which extends in cosmic scale the theory proposed by Ingraham\cite{ingr}, but with a different physical interpretation. In this theory we have another universal constant, $r_0$, which can be taken as $r/r_0 = N$, where 
$r$ is the radius of the hypersphere and $N$ is the cosmological number appearing in the  Eddington-Dirac theory\cite{eddi}.

Here we consider only the second alternative, i.e., the biprojective interpretation. With this 
aim we introduce a projective space $P_5$ tangent to the hypersphere $S^4$. We then introduce six projective coordinates ${\overline x}_a$, with $a=0,1,\ldots,5$ and normalized 
as 
$$
{\overline x}^2 + {\overline x}^2_0 - {\overline x}^2_ 5 = r^2,
$$
where ${\overline x}^2 = x_i x^i$, $i=1,\ldots,4$ and $r$ is the radius of the hypersphere. These coordinates allow us to construct the conformal projective relativity, using a six-dimensional tensor formalism.

This paper is organized as follows: in Section 1 we present a review of the so-called theory of hyperspherical universes, proposed by Arcidiacono, considering only the six-dimensional 
case, in which conformal projective relativity appears as a particular case. The choice 
of convenient coordinates and the link between the derivatives in these two formulations (a geometric version, six-dimensional, and a physical version, the five-dimensional conformal version) are also presented. After this review, we discuss in Section 2 a Klein-Gordon partial differential equation written in conformal coordinates. In Section 3, we show that a conformal `radial' d'Alembert-like equation, can be led 
into a Schr\"odinger differential equation in which the associated potential is exactly a second P\"oschl-Teller potential.

\section{Hyperspherical Universes}

In 1952 Fantappi\'e proposed the so-called theory of the hyperspherical universes\footnote{See the Appendix A.}.
This theory is based on group theory and on the hypothesis that the universe is endowed with unique physical laws, valid for all observers. As particular cases, Arcidiacono\cite{arci1} 
studied a limitation of that theory, i.e., he considered hyperspherical universes with 
$3,4,\ldots,n$ dimensions where motions are given by $n(n+1)/2$-parameter rotation group in spaces with $4,5,\ldots,{\mbox{(}}n+1{\mbox{)}}$ dimensions, respectively. Those models of hyperspherical $S^3, S^4,\ldots,S^n$ universes, can be interpreted 
as successive physical improvements, because any one of them (after $S^4$) contains its precedents and is contained in its successors.

After 1955 Arcidiacono studied the case $n=4$, special projective relativity, based on the 
de Sitter hyperspherical universe with a group (the so-called Fantappi\'e-de Sitter group) of ten parameters. This theory is an improvement (in a unique way) of Einstein's special relativity theory and provides a new group-theoretical version of the big-bang cosmology. As a by-product of special projective relativity one can recover several results, for example, Kinematic Relativity, proposed by Milne\cite{miln}; Stationary Cosmology, proposed by Bondi-Gold\cite{bond} and Plasma Cosmology, proposed by Alfv\`en\cite{alfv}.  

Moreover, if we consider a universe $S^4$ as globally hyperspherical but endowed with a locally variable curvature, we obtain the so-called general projective relativity which was proposed and studied by Arcidiacono after 1964. This theory allows us to recover several results as particular cases, for example, the unitary theories proposed by Weyl\cite{weyl}, 
Straneo\cite{stra}, Kaluza-Klein\cite{kalu,klei}, Veblen\cite{vebl} and Jordan-Thiry\cite{jord}
and some generalizations of the gravitational field, as those proposed by Brans-Dicke\cite{bran}, Rosen\cite{rose} and Sciama\cite{scia}.

In this paper we are interested only in the case $n=5$, i.e., conformal projective relativity based on the hyperspherical universe $S^4$ and its associated rotation group, with fifteen parameters, which contains the accelerated motions. We remember that, whereas for $n=4$ we have a unitary theory (a magnetohydrodynamic field), for $n=5$ we have another unitary theory, i.e.,
the magnetohydrodynamics and Newton's gravitation. We also present the relations between Cartesian, projective and conformal coordinates and the link involving derivatives in the six- and five-dimensional formulations.

\subsection{Conformal Coordinates}
We use the notation $x_i$, ($i=1,2,3,4$) and $x_5$ for conformal coordinates and 
${\overline x}_a$, ($a=0,1,2,3,4,5$) for projective coordinates. The relations between 
these coordinates are
$$
x_i = r_0 \frac{\overline{x}_i}{\overline{x}_0 + \overline{x}_5}
\qquad {\mbox{and}} \qquad x_5 = r_0 \frac{r}{\overline{x}_0 + \overline{x}_5},
$$
which satisfy the condition
$$
x_5^2 - x^2 = r_0^2 \frac{\overline{x}_0 - \overline{x}_5}
{\overline{x}_0 + \overline{x}_5},
$$
where $x^2 = x_i x^i$, and $r_0$ and $r$ are constants. After these considerations, the transformations of the so-called conformal projective group are obtained using the quadratic form in projective coordinates
$$
{\overline x}^2 + {\overline x}^2_0 - {\overline x}^2_ 5 = r^2,
$$
decomposing the elements of the six-dimensional rotation group (with fifteen parameters) in fifteen simple rotations (${\overline x}_a, {\overline x}_b$).
 
\subsection{Connection Between Derivatives}
Our main objective is to write down a differential equation, more precisely a Klein-Gordon-like equation, associated with conformal coordinates. We first obtain the relation 
between the six projective derivatives ${\overline \partial}_a \equiv \partial / \partial {\overline x}_a$ and the five-dimensional derivatives $\partial_i = \partial / \partial x_i$ 
and $\partial_5 = \partial / \partial x_5$. We can then write the differential 
equations in the projective formalism, with six dimensions, in physical, i.e., conformal coordinates, with five dimensions.\footnote{As we already know, in five dimensions 
we must impose a condition on space in order to account for the fact that we are aware of only four dimensions. We have the same situation here, i.e., we must impose an additional condition.}

Taking $\phi = \phi (x_i , x_5)$, a scalar field, and using the chain rule we can write
$$
\begin{array}{ccl}
\partial_i \phi &=& \left[ ({\overline\partial}_i {\overline x}_k){\overline \partial}_k + ({\overline \partial}_i {\overline x}_5){\overline \partial}_5 + ({\overline \partial}_i {\overline x}_0) {\overline \partial}_0 \right] {\overline \phi}\\
\partial_5 \phi &=& \left[({\overline \partial}_5 {\overline x}_k) {\overline \partial}_k +
({\overline \partial}_5 {\overline x}_5){\overline \partial}_5 + ({\overline \partial}_5 {\overline x}_0) {\overline \partial}_0 \right] {\overline \phi}
\end{array}
$$
with ${\overline \phi} = {\overline \phi} ({\overline x}_i , {\overline x}_5 , {\overline x}_0)$ and $ i,k=1,2,3,4.$

From now on we take $r=1=r_0$. We consider ${\overline \phi}({\overline x}_a)$ a 
homogeneous function with degree $N$ in all six projective coordinates ${\overline x}_a$. 
Using Euler's theorem associated with homogeneous function, we get
$$
\left( \overline{x}_i \overline{\partial}_i + \overline{x}_5 \overline{\partial}_5 +
\overline{x}_0 \overline{\partial}_0 \right) \phi = N \phi
$$
where $\overline{\partial}_a = \partial / \partial \overline{x}_a$ and $N$ is the degree of homogeneity of the function. 

Then, the link between the derivatives can be written as follows\cite{arci3}
$$
\begin{array}{ccl}
\overline{\partial_0} \,\overline{\phi} &=& N\displaystyle \frac{A^{+}}{x_5} \phi +
B^{-} \partial_5 \phi - x_5 x_i \partial_i \phi\\
&&\\
\overline{\partial_5} \,\overline{\phi} &=& -N\displaystyle \frac{A^{-}}{x_5} \phi
- B^{+} \partial_5 \phi - x_5 x_i \partial_i \phi\\
&&\\
\overline{\partial_i} \,\overline{\phi} &=& N\displaystyle \frac{x_i}{x_5} \phi +
x_i \partial_5 \phi +  x_5 \partial_i \phi
\end{array}
$$
where we have introduced a convenient notation
$$
2A^{\pm} = 1 \mp x^2 \pm x_5^2 \qquad {\mbox{and}} \qquad  2B^{\pm} = 1\pm x^2 \pm x_5^2.
$$
We observe that for ${\overline x}_5 = 0$ and considering ${\overline \partial}_5 {\overline \phi} = 0$ we obtain
$$
\begin{array}{ccl}
\overline{\partial}_i \overline{\phi} & = & A \partial_i \phi +
\displaystyle \frac{N}{A} x_i \phi\\
&&\\
\overline{\partial}_0 \overline{\phi} & = & - A x_i \partial_i
\phi + \displaystyle \frac{N}{A} \phi
\end{array}
$$
where $A^2 = 1 + x^2$. These expressions are the same expressions obtained in special projective relativity\cite{eco3} and provide the link between the five projective derivatives and the four derivatives in Cartesian coordinates, i.e., the relation between five-dimensional (de Sitter) universe and four-dimensional (Minkowski) universe.

\section{Conformal Klein-Gordon Equation}
In this section we use the  previous results to calculate the so-called generalized Klein-Gordon differential equation 
$$
\frac{\partial^2}{\partial \overline{x}_a^2} \Phi + {\sf m }^2
\Phi = 0
$$
where ${\sf m}^2$ is a constant and $a=0,1,\ldots ,5.$ Introducing projective coordinates 
(in this case we have a Pythagorean metric) we obtain\footnote{Hereafter we consider 
${\sf m} = m_0 c / \hbar$ where $m_0$, $c$ and $\hbar$ have the usual meanings.} 
$$
\frac{\partial^2 U}{\partial \overline{x}_i^2} + \frac{\partial^2 U}{\partial
\overline{x}_0^2}- \frac{\partial^2 U}{\partial \overline{x}_5^2}
+ {\sf m}^2 U = 0
$$
where $i=1,2,3,4$ and $U=U(\overline{x}_i , \overline{x}_0 , \overline{x}_5)$. 

Using the relations between projective and conformal coordinates 
and the link (involving the derivatives) in the two formulations we can write
$$
\left[ x_5^2 \left( \Box - \frac{\partial^2}{\partial x_5^2}
\right) + 3x_5 \frac{\partial}{\partial x_5} + N(N+5) + {\sf
m}^2 \right] u(x_i , x_5) = 0
$$
where $N$ and ${\sf m}^2$ are constants, $\Box$ is the Dalembertian operator given by
$$
\Box = \Delta - \frac{1}{c^2} \frac{\partial^2}{\partial t^2}
$$
and $\Delta$ is the Laplacian operator. This partial differential equation is the 
so-called Klein-Gordon differential equation written in conformal coordinates or 
a conformal Klein-Gordon equation.

The case ${\sf m}^2 = 0$ transforms this equation in the so-called generalized d'Alembert differential equation. Another way to obtain this differential equation is to consider 
the conformal metric in Cartesian coordinates, which furnishes the so-called Beltrami metric\cite{arci1} where the d'Alembert equation appears naturally. This equation 
can also be obtained by means of the second order Casimir invariant operator\footnote{Invariant operators associated with dynamic groups furnish mass formulas, energy spectra and, in general, characterize specific properties of physical systems.} associated with the conformal group.

To solve the conformal Klein-Gordon equation, we first introduce the spherical coordinates 
$(r, \theta , \phi)$ and get
\bege
\frac{\partial^2 u}{\partial r^2} + \frac{2}{r} \frac{\partial
u}{\partial r} + \frac{1}{r^2} {\cal L}u  - \frac{1}{c^2} \frac{\partial^2
u}{\partial t^2} - \frac{\partial^2 u}{\partial x_5^2} +
\frac{3}{x_5} \frac{\partial u}{\partial x_5} +
\frac{\Lambda}{x_5^2} u = 0,\label{201}
\enge
where we introduced $x_4 = i ct$ and defined the operator\footnote{Here $r$ is a coordinate and should not be confused with the radius of the hypersphere. Besides, it is always possible to define a Wick-rotation\cite{saku} of the time coordinate, i.e., $ct$ $\mapsto$ $ict$.} 
\bege
{\cal L} \equiv  \frac{\partial^2
}{\partial \theta^2} + \cot \theta \frac{\partial }{\partial
\theta} + \frac{1}{\sin^2 \theta} \frac{\partial^2
}{\partial \phi^2}
\enge
involving only the angular part. In this partial differential equation we have $u=u(r,\theta , \phi , t, x_5)$ with $\Lambda = N(N+5) + {\sf m}^2$.

Using the method of separation of variables we can eliminate the temporal and angular parts, writing 
\bege\label{200}
u=u(r,\theta , \phi , t, x_5) = A\, {\mbox{e}}^{inct} Y_{\ell m} (\theta , \phi) \, f(r, x_5),
\enge
where $A$ is an arbitrary constant, $n > 0$, $\ell = 0,1,\ldots$ and $m = 0, \pm 1, \ldots$ with $-\ell \leq m \leq \ell$ and $Y_{\ell m} (\theta , \phi)$ are the spherical harmonics, we get 
the following partial differential equation
$$
\frac{\partial^2 f}{\partial r^2} + \frac{2}{r} \frac{\partial
f}{\partial r} - \frac{\partial^2 f}{\partial x_5^2} +
\frac{3}{x_5} \frac{\partial f}{\partial x_5} +
\frac{\Lambda}{x_5^2} f + \left[n^2 - \frac{\ell (\ell +1)}{r^2}
\right] f = 0
$$
with $f=f(r,x_5)$. If we impose a regular solution at the origin ($r \to 0$), the solution 
of this partial differential equation can be obtained in terms of a product of two Bessel 
functions.

\section{A d'Alembert-Like Equation}
In this section we present and discuss a partial differential equation which can be 
identified to a d'Alembert-like equation, which we call a conformal `radial' d'Alembert equation. We firstly introduce a convenient new set of coordinates, then we use separation of variables and obtain two ordinary differential equations. One of them can be identified as an ordinary differential equation whose solution is a generalization of Newton's law of gravitation; the other one is identified with an ordinary differential equation similar to a one-dimensional Schr\"odinger differential equation with a potential equal to the second P\"oschl-Teller potential.  

We introduce the following change of independent variables 

\begin{eqnarray}
r & = & \rho \cosh \xi,\nonumber\\
x_5 & = & \rho \sinh \xi,\label{203}
\end{eqnarray}
\noi 
with $\rho >0$ and $\xi \geq 0$, in the separated Klein-Gordon equation, obtained in the previous section, and after another separation of variables we can write a pair of ordinary differential equations, namely,  
\bege
\rho^2 \frac{d^2 U}{d\rho^2} - p(p+1)U = 0,\label{204}
\enge
where $U=U(\rho)$ and
\bege
\frac{d^2 V}{d\xi^2} + (2 \tanh \xi - 3 \coth \xi)\frac{dV}{d\xi}
+ \left[ \frac{\ell (\ell +1)}{\cosh^2 \xi} -
\frac{\Lambda}{\sinh^2 \xi} - p(p+1)\right] V =0\label{205}
\enge
where $V=V(\xi)$ and $p$ is a separation constant.

We first discuss eq.(\ref{204}). Its general solution is given by
$$
U(\rho) = C_1 \rho^{\,-p} + C_2 \rho^{\, p+1}
$$
where $C_1$ and $C_2$ are arbitrary constants.

If we consider the case $p=1$, introducing the notation $C_1 = gM$ with $g$ and $M$ having the usual meanings, we get
$$
U(\rho) = \frac{gM}{\rho} + C_2 \, \rho^2
$$
i.e., a gravitational potential which can be interpreted as a sum of a Kepler-like potential and a harmonic oscillator potential, giving rise to the gravitational force
$$
f(\rho) = \nabla U = - \frac{gM}{x^2 - x_5^2} + 2C_2(x^2 -
x_5^2)^{1/2},
$$
with a singularity at $x=x_5$. We note that for $C_2 = 0$ we obtain an expression analogous to  Newton's law of gravitation.

Secondly, eq.(\ref{205}) can be solved by introducing the change of dependent variable
$$
V(\xi) = \sinh^{\frac{1}{2}} \xi  \tanh \xi \, F(\xi)
$$ 
and we obtain
\begin{equation}\label{14}
-\frac{d^2}{d\xi^2}F(\xi) + \left[ \frac{\mu (\mu - 1)}{\sinh^2 \xi} - \frac{\ell (\ell+1)}{\cosh^2 \xi} + \left(p+ \frac{1}{2}\right)^2 \right] F(\xi ) = 0,
\end{equation}
where the parameter $\mu$ is given by a root of the algebraic equation $\mu (\mu - 1) = N^2 + 5N + 15/4 + {\sf m}^2$.


The differential equation above can be identified with a Schr\"odinger-like differential 
equation in which the associated potential is given by
\begin{equation}\label{209}
{\cal V}_{\mu \ell} (\xi) = \frac{\mu (\mu - 1)}{\sinh^2 \xi} - \frac{\ell (\ell+1)}{\cosh^2 \xi},
\end{equation}
which is exactly the second P\"oschl-Teller potential with energy $E$ given by $E_p = -(p+1/2)^2 < 0$. 
The graphics showing $V_{\mu\ell}(\xi)$ for various values of $\mu$ and $\ell$ are depicted below:

\begin{figure}[H]
\centering
\includegraphics[width=8.7cm]{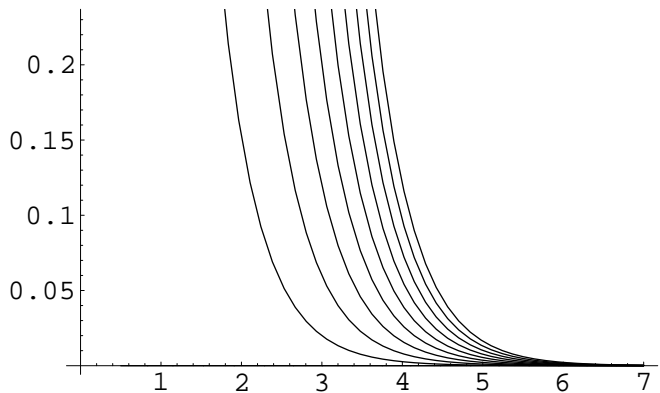}
  \caption{\small $V_{\mu\ell}(\xi)\times \xi$ evaluated for $\mu = \{1,\ldots,10\}$ and $\ell = 0,1$.}
\label{n1}
\end{figure}

\begin{figure}[H]
\centering
\includegraphics[width=8.7cm]{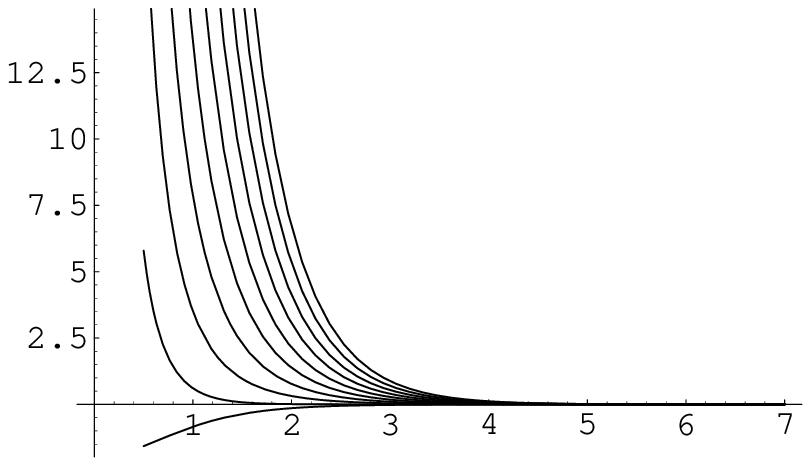}
  \caption{\small $V_{\mu\ell}(\xi)\times \xi$ evaluated for $\mu = \{1,\ldots,10\}$ and $\ell = 2$.}
\label{n1}
\end{figure}

\begin{figure}[H]
\centering
\includegraphics[width=8.7cm]{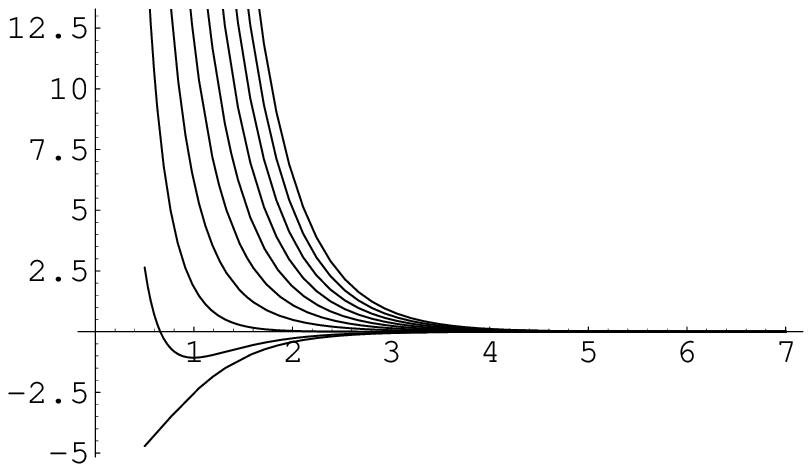}
  \caption{\small $V_{\mu\ell}(\xi)\times \xi$ evaluated for $\mu = \{1,\ldots,10\}$ and $\ell = 3$.}
\label{n1}
\end{figure}

\begin{figure}[H]
\centering
\includegraphics[width=8.7cm]{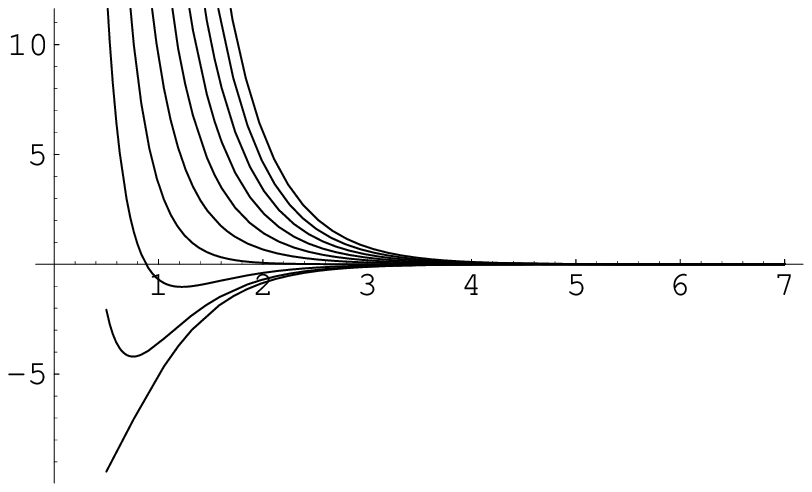}
  \caption{\small $V_{\mu\ell}(\xi)\times \xi$ evaluated for $\mu = \{1,\ldots,10\}$ and $\ell = 4$.}
\label{n1}
\end{figure}

\begin{figure}[H]
\centering
\includegraphics[width=8.7cm]{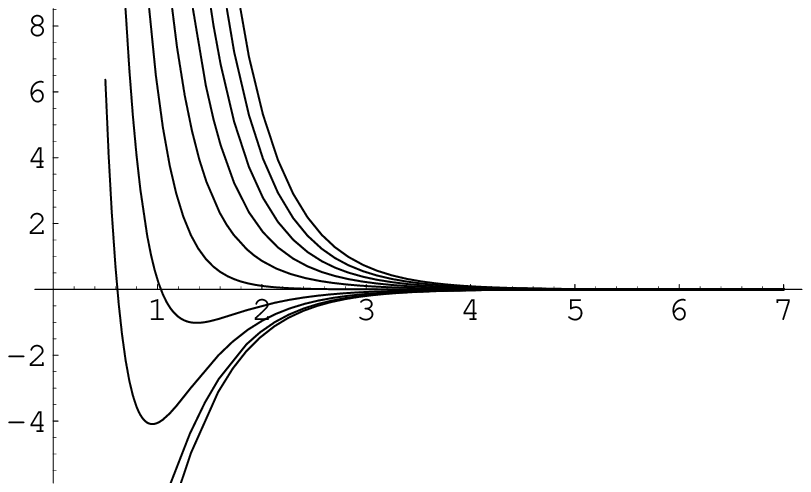}
  \caption{\small $V_{\mu\ell}(\xi)\times \xi$ evaluated for $\mu = \{1,\ldots,10\}$ and $\ell = 5$.}
\label{n1}
\end{figure}

\begin{figure}[H]
\centering
\includegraphics[width=8.7cm]{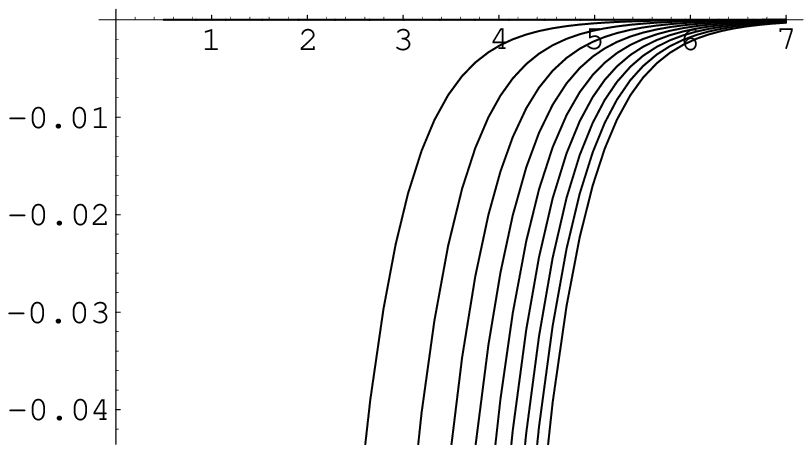}
  \caption{\small $V_{\mu\ell}(\xi)\times \xi$ evaluated for $\ell = \{1,\ldots,10\}$ and $\mu = 0,1.$}
\label{n1}
\end{figure}

\begin{figure}[H]
\centering
\includegraphics[width=8.7cm]{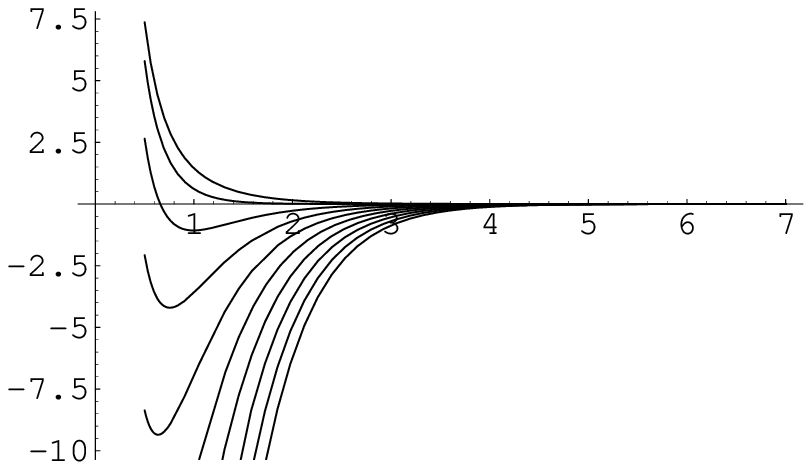}
  \caption{\small $V_{\mu\ell}(\xi)\times \xi$ evaluated for $\ell = \{1,\ldots,10\}$ and $\mu = 2.$}
\label{n1}
\end{figure}

\begin{figure}[H]
\centering
\includegraphics[width=8.7cm]{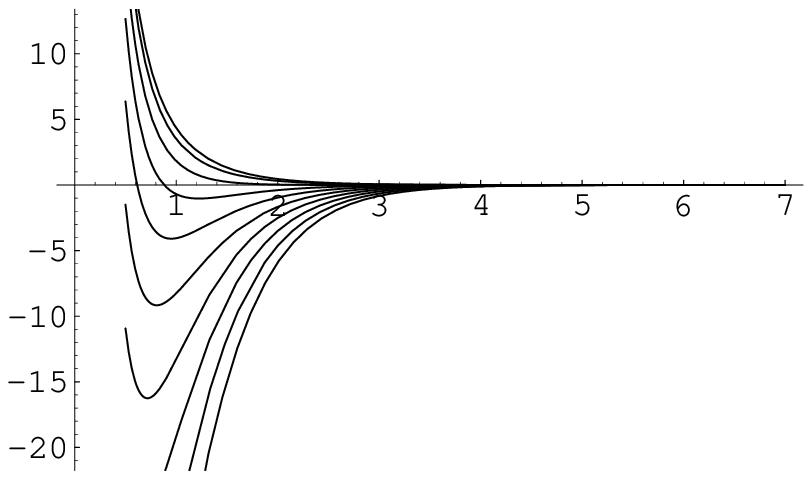}
  \caption{\small $V_{\mu\ell}(\xi)\times \xi$ evaluated for $\ell = \{1,\ldots,10\}$ and $\mu = 3.$}
\label{n1}
\end{figure}

\begin{figure}[H]
\centering
\includegraphics[width=8.7cm]{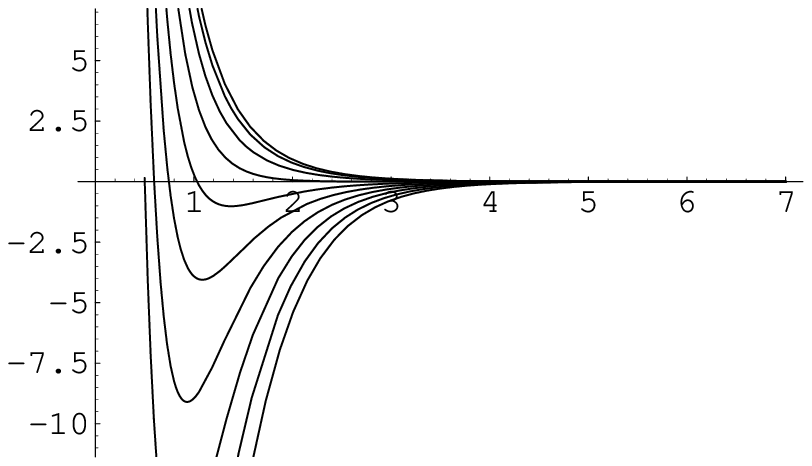}
  \caption{\small $V_{\mu\ell}(\xi)\times \xi$ evaluated for $\ell = \{1,\ldots,10\}$ and $\mu = 4.$}
\label{n1}
\end{figure}

\begin{figure}[H]
\centering
\includegraphics[width=8.7cm]{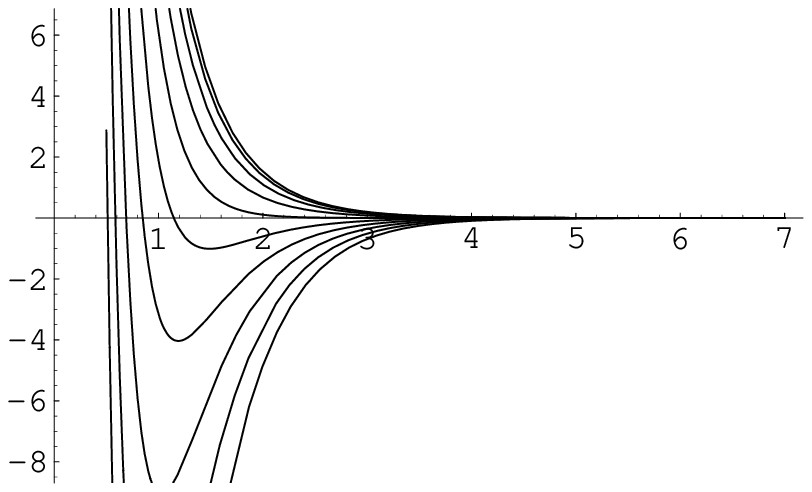}
  \caption{\small $V_{\mu\ell}(\xi)\times \xi$ evaluated for $\ell = \{1,\ldots,10\}$ and $\mu = 5.$}
\label{n1}
\end{figure}

 Explicit solutions of eq.(\ref{14}) are written as
\begin{eqnarray}
F(\xi) &=& (\tanh^{-1/2}\xi)(\tanh^2\xi-1)^\sigma\times\nonumber\\
&&\times\left( {}_2F_1(\sigma_1,\sigma_2,\sigma_3,\tanh^2\xi)\,(\tanh^{-\sigma_3}\xi)D + 
 {}_2F_1(1+\sigma_3 + \sigma_1,1+\sigma_2+\sigma_3,2+\sigma_3,
\tanh^2\xi)\,(\tanh^{2+\sigma_3}\xi) E\right)\nonumber
\end{eqnarray}
\noindent where $D, E$ are integration constants, ${}_2F_1$ denotes the hipergeometric function, and
\begin{equation}
\sigma_1 = \frac{1}{4}(3-2\mu-2\ell-2p),\quad \sigma_2 = \frac{1}{4}(1-2\mu+2\ell-2p),\quad 
\sigma_3 = \frac{1}{2}(2\mu-3),\quad 2\sigma = \sigma_1 + \sigma_2 + \sigma_3\nonumber
\end{equation}\noindent


We note that the first P\"oschl-Teller potential is connected with the 
study of a Dirac particle on central backgrounds associated with an anti-de Sitter 
oscillator, i.e., the transformed radial wave functions satisfy the second-order 
Schr\"odinger differential equation whose potential is exactly the first P\"oschl-Teller
potential\cite{cota}.

\section{QNMs equations}
It is also possible to derive QNMs formal partial differential equations if, instead 
the \emph{ansatz} given by eq.(\ref{200}), we assume
\bege
u=u(r,\theta,\phi,t,x_5) = e^{\pm im\phi}\,F(r,\theta,t,x_5),
\enge\noi  
and eq.(\ref{201}) reads
\bege
\left(\frac{\partial^2}{\partial r^2} + \frac{2}{r} \frac{\partial
}{\partial r} + \frac{1}{r^2} \frac{\partial^2
}{\partial \theta^2} + \cot \theta \frac{\partial }{\partial
\theta} + \frac{1}{\sin^2 \theta} \frac{m^2}{r^2\sin^2\theta}  - \frac{1}{c^2} \frac{\partial^2
}{\partial t^2} - \frac{\partial^2}{\partial x_5^2} +
\frac{3}{x_5} \frac{\partial}{\partial x_5} +
\frac{\Lambda}{x_5^2}\right)F(r,\theta,t,x_5) = 0,
\enge\noi which, after the separation of variables given by $F(r,\theta,t,x_5) = \Theta(\theta)R(r,t,x_5)$,  can be lead to 
\bege
\Theta''(\theta) + \cot\theta\Theta'(\theta) +[\ell(\ell+1)-m^2\csc^2\theta]\Theta(\theta) = 0,
\enge\noi which presents the associated Legendre polynomials $\Theta(\theta) = P_\ell^m(\theta)$ as solutions \footnote{Here we have imposed regularity 
condition, i.e., for $\theta = \pm \pi$, the solution is analytic.}, and 
\bege
\left(\frac{\partial^2}{\partial r^2} + \frac{2}{r} \frac{\partial
}{\partial r}  - \frac{1}{c^2} \frac{\partial^2
}{\partial t^2} - \frac{\partial^2}{\partial x_5^2} +
\frac{3}{x_5} \frac{\partial}{\partial x_5} +
\left(\frac{\Lambda}{x_5^2}-\frac{\ell(\ell+1)}{r^2}\right)\right)R(r,t,x_5) = 0,\label{207}
\enge\noi
Now if we use again eqs.(\ref{203}), then eq.(\ref{207}) takes the form
\bege
-\frac{\partial^2 R}{\partial t^2} + \frac{\partial^2 R}{\partial \rho^2}
-\frac{1}{\rho^2}\left[\frac{\partial^2}{\partial\xi^2} + (2 \tanh \xi - 3 \coth \xi)\frac{\partial}{\partial\xi}
+ \left( \frac{\ell (\ell +1)}{\cosh^2 \xi} -
\frac{\Lambda}{\sinh^2 \xi}\right)\right]R(\xi,t,\rho) =0\label{208}
\enge
where now $R=R(\xi,t,x_5)$. By means of the \emph{ansatz}
\bege
R(\xi,t,x_5) = \frac{\sinh^{3/2}(\xi)}{\cosh(\xi)}H(\xi,t,x_5),
\enge and making $\rho = 1$, which implies by eqs.(\ref{203}) that $r^2 - x_5^2 = 1$,  eq.(\ref{208}) gives
\begin{equation}\label{210}
\frac{\partial^2}{\partial t^2} + \left[\frac{\partial^2}{\partial\xi^2} - V_{\mu,\ell}(\xi) -\frac{1}{4} \right] u_1(\xi,t) = 0,
\end{equation} \noi where $u_1(\xi,t) = R(\xi,t,\rho=1)$. The potential $V_{\mu,\ell}(\xi)$ is defined by eq.(\ref{14}), 
where $\mu(\mu - 1) = \Lambda + 15/4$.
These equations are formally QNMs equations and are prominently used in the description of gravitational 
and electromagnetic perturbations in supermassive stellar and black holes processes.

\section*{Concluding Remarks}
In this paper we discussed the calculation of a conformal d'Alembert-like equation.
We used the methodology of projective relativity to obtain a conformal Klein-Gordon 
differential equation and, after the separation of variables, we got another partial differential equation in
 only two independent variables, the so-called conformal d'Alembert differential equation. Another separation 
of variables led to an ordinary differential equation which generalizes Newton's law of gravitation. Finally,
 we have shown that the remaining differential equation, a `radial' differential equation, is transformed into a 
one-dimensional Schr\"odinger differential equation with an associated potential that can be identified exactly
 with the second P\"oschl-Teller potential.  These equations can shed some new light on the analytical investigations
concerning QNMs.
There are some open questions concerning QNM modes, and new developments are promising, and not only restricted to $d=4$ spacetimes. 
Motl and Neitzke \cite{20}
 has recently given an original derivation of the quasinormal frequencies of Schwarzschild black holes in the case 
where $d\geq 4$, and also of Reissner-Nordstr\o m black holes in $d = 4$, in the limit of infinite damping, where they 
find an aperiodic behavior for the QNM for spin 0,1, and 2 fields \cite{20}. Also, in  \cite{21,22} QNMs are investigated 
in  a time-dependent black hole background and in Reissner-Nordstr\o m-AdS black holes. 

From supersymmetric quantum mechanics with periodic potentials, it can be seen that the most general periodic potentials
 which can be analytically solved involve Jacobi's 
elliptic functions, which in various limits become P\"oschl-Teller potentials arising in the context of Kaluza-Klein spectrum\cite{kalu}. Kaluza-Klein modes of the graviton have been widely investigated \cite{randall,hata,bram,nam}, since the original formulation of Randall and Sundrum necessarily has a continuum of Kaluza-Klein modes without any mass gap, arising from a periodic system of 3-branes. The methods and equations developed here can shed some new light in the calculation of mass gaps from a distribution of $D$-branes\cite{nam} in the context of  five-dimensional supergravity -- which will be discussed in a forthcoming paper.

A natural continuation of this calculation is to prove that all `radial' problems associated with an equation resulting from a problem involving a light cone can be led into a Schr\"odinger-like differential equation in which the potential is exactly the P\"oschl-Teller potential\cite{ECO}.

\section*{Acknowledgment}
We are grateful to Dr. Geraldo Agosti for interesting and useful discussions and 
Prof. Waldyr A. Rodrigues Jr., Dr. J. Em\'{\i}lio Maiorino for several suggestions about the paper. 

\appendix

\section{Hyperspherical Universe Models}

In this appendix we briefly summarize the idea of hyperspherical universes as
originally proposed by Fantappi\'{e} \cite{fant} and developed by Arcidiacono
\cite{arci1}.

The main motivation behind those models is to consider seriously the premise that
a Universe must be a harmonic and well ordered system of laws, and that this
statement is to be expressed mathematically by using group theory in an
appropriate way. Taking into account that Galilean Relativity, which uses the

Galileo group as invariance group of physical laws, has been perfectioned
into Special Relativity, which uses the Poincar\'{e} group as invariance group
of physical laws, Fantappi\'{e} asked himself in which way it would be
possible to perfect Special Relativity into a kind of \textit{final} relativity.
His answer to this question was very simple indeed. He realized that the
Poincar\'{e} group is the contraction of the 10-parameter Lie group known today as
de Sitter group (but which should be called, as in the text, the 
Fantappi\'{e}-de Sitter group) which can be made to act projectively on a flat
4-dimensional (Minkowski) spacetime. Next Fantappi\'{e}
asked whether there were other spacetime manifolds where the group could act
naturally and serve as invariance group of physical laws. The answer is
positive, and Fantappi\'{e} found that the natural manifold is a
hyperspherical universe, which he called $S^{4}$, of constant curvature radius.
Of course, the previously known models for the universe (not based on General
Relativity) are all particular cases of this one, corresponding to some group
contractions involving the velocity of light and/or the universe radius, or
both. Fantappi\'{e} did not stop there. He proposed that the hyperspherical
universe model $S^{4}$ was only an approximation for truth in the sense that
it was embedded in a hyperspherical universe model $S^{5}$ where the
conformal group (a 15-parameter Lie group) acts naturally as invariance group
of physical laws. By its turn, $S^{5}$ may be generalized into $S^{6}$ and so
on. At each generalization new fundamental physical constants make their
natural appearance as a kind of group parameter whose contraction produces the
group used in the previous universe model. It is clear that the extra
dimensions in each Universe model must be interpreted, in an appropriate way
(something that is also necessary in modern Kaluza-Klein type theories), and it is
at this point that his mathematical skills give us useful physical hints. In
particular, Arcidiacono, one of his students, showed that the hyperspherical
universe models $S^{4}$ and $S^{5}$ contain many aspects of several proposed
unified theories. Of course, we do not have space here even to start
discussing the many beautiful results found by Arcidiacono and we invite the
reader to consult Arcidiacono book\cite{arci1} for more details. However, we would like to emphasize here that many ideas proposed by him are worth to be more developed since, in
particular, it seems that in his work there is the seed of a simple
solution for the problem of dark energy and dark matter, an issue that we
shall discuss elsewhere.

\end{document}